\documentclass[bibyear]{aa}
\usepackage[varg]{txfonts}
\usepackage{graphicx}
\usepackage{subfigure}
\usepackage{amsmath}
\usepackage{amssymb}
\usepackage{natbib}
\usepackage{multirow}
\usepackage{longtable}
\bibpunct{(}{)}{;}{a}{}{,}
\newcommand{\changefont}[3]{
\fontfamily{#1}\fontseries{#2}\fontshape{#3}\selectfont}

\begin{document}

\titlerunning{Radio emission of young solar-type stars}
\authorrunning{Fichtinger et al.}

\title{Radio emission and mass loss rate limits of four young solar-type stars}
\author{Bibiana Fichtinger \inst{1} \and Manuel Güdel \inst{1} \and Robert L. Mutel \inst{2} \and Gregg Hallinan \inst{3} \and Eric Gaidos \inst{4} \and Stephen L. Skinner \inst{5} \and Christene Lynch \inst{2} \and Kenneth G. Gayley \inst{2}}
\offprints{B. Fichtinger, \email{bibiana.fichtinger@univie.ac.at}}
\institute{Institute of Astrophysics, University of Vienna, T\"urkenschanzstrasse 17, 1180 Vienna, Austria \and Department of Physics and Astronomy, University of Iowa, 203 Van Allen Hall, Iowa City, IA 52242, USA \and Department of Astronomy, California Institute of Technology, 1200 E. California Blvd., Pasadena, CA 91125, USA \and Department of Geology and Geophysics, University of Hawaii, Honolulu, Hawaii, HI 96822, USA \and Center for Astrophysics and Space Astronomy, University of Colorado, Boulder, CO 80309-0389, USA}
\date{Received date / Accepted date}

\abstract{} {Observations of free-free continuum radio emission of four young main-sequence solar-type stars (EK Dra, $ \rm \pi^{1} \ UMa $, $ \rm \chi^{1}\  Ori, $ and $ \rm \kappa^{1} \ Cet) $ are studied to detect stellar winds or at least to place upper limits on their thermal radio emission, which is dominated by the ionized wind. The stars in our sample are members of The Sun in Time programme and cover ages of $ \sim $ 0.1 - 0.65 Gyr on the main-sequence. They are similar in magnetic activity to the Sun and thus are excellent proxies for representing the young Sun. Upper limits on mass loss rates for this sample of stars are calculated using their observational radio emission. Our aim is to re-examine the faint young Sun paradox by assuming that the young Sun was more massive in its past, and hence to find a possible solution for this famous problem.} {The observations of our sample are performed with the Karl G. Jansky Very Large Array (VLA) with excellent sensitivity, using the C-band receiver from 4 - 8 GHz and the Ku-band from 12 - 18 GHz. Atacama Large Millimeter/Submillitmeter Array (ALMA) observations are performed at 100 GHz. The Common Astronomy Software Application (CASA) package is used for the data preparation, reduction, calibration, and imaging. For the estimation of the mass loss limits, spherically symmetric winds and stationary, anisotropic, ionized winds are assumed. We compare our results to 1) mass loss rate estimates of theoretical rotational evolution models, and 2) to results of the indirect technique of determining mass loss rates: Lyman-$\alpha$ absorption.} {We are able to derive the most stringent direct upper limits on mass loss so far from radio observations. Two objects, EK Dra and $ \rm \chi^{1}\  Ori $, are detected at 6 and 14 GHz down to an excellent noise level. These stars are very active and additional radio emission identified as non-thermal emission was detected, but limits for the mass loss rates of these objects are still derived. The emission of $ \rm \chi^{1}\  Ori $ does not come from the main target itself, but from its M-dwarf companion. The stars $ \rm \pi^{1} \ UMa $ and $ \rm \kappa^{1} \ Cet $ were not detected in either C-band or in Ku-band. For these objects we give upper limits to their radio free-free emission and calculate upper limits to their mass loss rates. Finally, we reproduce the evolution of the Sun and derive an estimate for the solar mass of the Sun at a younger age.} {}

\keywords{Sun: evolution - stars: winds, outflows - stars: solar-type - stars: mass loss}
\maketitle

\section{Introduction}
\label{sec:intro}
Geological evidence suggests that the early Earth had a warmer climate in the first few 100 Myr of its evolution. Such a mild and warm climate on the early Earth 4 Gyr ago was necessary and essential for the evolution and formation of life on our planet \citep{Kasting1989, Sackmann2003}. However, solar standard models predict a lower bolometric luminosity of the Sun at that time, being just 70\% the present-day luminosity. The evolution of the Sun's luminosity had an important effect on the formation of the atmosphere for our Earth and for early Mars. Without an atmosphere on Earth, the average surface temperature would have been 235 K only. Additional present-day greenhouse gases would have raised the temperature to $\sim$ 253 K, which is still not enough to avoid the completely frozen surfaces on early Earth and Mars \citep{Sagan1972, Kasting2003}. The discrepancy between the implications from the solar standard models and the geological evidence for a warmer climate on Earth is defined as the “faint young Sun paradox” (FYSP). Apart from a number of proposed solutions of the FYSP \citep[see e.g.][]{Gaidos2000, Feulner2012}, an astrophysical solution for this problem has been suggested. It assumes that the young main-sequence Sun was brighter than suggested by the standard model, which would be possible if it had been more massive than today and consequently suffered from an increased mass loss during its early main-sequence life through an enhanced solar wind \citep{Graedel1991, Gaidos2000}. 

Winds play an important role in stellar evolution for main-sequence stars like the Sun, especially for the stellar angular momentum. We know that stars spin down with age, because angular momentum is carried away by the magnetized, ionized winds. To understand the mechanism of the interaction between the stellar wind, stellar rotation, and the magnetic field for stars with various ages, information on how winds evolve with time is required. Furthermore, the evolution of stellar winds is important for the evolution of planetary atmospheres and their erosion \citep{Lammer2010}. Most of what we know about stellar winds comes from studies of the solar wind, although the mechanisms for generating, accelerating, and heating the solar wind are still poorly understood \citep{Cranmer2009, McComas2003, Schwenn2006}.\\

Today, the most common way to assess stellar winds and therefore, determine stellar mass loss rates, is to observe the Lyman-$\alpha$ excess of the neutral interstellar hydrogen in high-resolution Hubble Space Telescope spectra of stars, as introduced by \citet{Wood2002}, \citet{Wood2004}, and \citet{Wood2005a}. Due to charge exchange interactions between neutral interstellar hydrogen and the ionized wind, a "wall" of hot neutral hydrogen at the edge of the stellar astrosphere is built up. The detected material is not from the fully ionized wind itself, which has no HI, but is interstellar HI instead that is heated within the interaction region between the wind and the local interstellar medium. The amount of astrospheric HI absorption provides diagnostic information on the rate of mass loss of the wind, specifically the momentum in the wind. In \citet{Wood2002} the correlation of the mass loss rates with coronal properties was studied. The coronal X-ray luminosity is a good indicator for the magnetic activity of a star and the scaling relationship between the mass loss rate per unit surface area and the X-ray surface flux is

\begin{equation}
\dot{M} \propto F_{X}^{1.34 \ \pm \ 0.18},
\label{eq:wood_flux}
\end{equation}

\noindent
which, combined with X-ray luminosity evolution versus time $ L_{X} \propto t^{-1.5} $ \citep{Guedel1997}, suggests that the mass loss rate decreases with time for solar-like stars like $ \dot{M} \propto t^{-2.33\pm0.55} $ \citep{Wood2004}. The correlation is, however, still not sufficient to solve the FYSP \citep{Wood2002, Minton2007}. However, as these authors concluded in their study, this correlation fails for the youngest and most active stars for which winds appear to be very weak in Lyman-$\alpha$ observations. \citet{Wood2005b} observed $ \rm \chi^{1} \ Ori $, but they were unable to provide any astrospheric detections in their study. They argued that a non-detection does not generally provide a meaningful upper limit to the stellar wind strength, because for non-detections the star could be surrounded by a fully ionized interstellar medium (ISM) \citep{Wood2005b}. \\

Observing and detecting stellar winds similar to the solar case is important to improve our understanding of stellar evolution, such as the correlation between rotation and stellar magnetic activity which provides information on the dynamo and thus magnetic activity. Furthermore, the understanding of acceleration mechanisms of these winds could be improved and the measurements of wind properties of stars with different ages may provide essential information on stellar angular momentum loss. Radio observations of young, solar-type stars are used in our study to test if there was a strong mass loss in the young Sun. A study of the "radio Sun in time", complementing the "X-ray Sun in time" \citep{Guedel1997}, can explore the range and the long-term evolution of solar and stellar magnetic activity and wind mass loss. First detections of radio emission from low-mass main-sequence stars were reported by \citet{Gary1981} and \citet{Linsky1983}. Limits to mass loss have already been established from radio observations of more massive A and F stars \citep{Brown1990} and active, less massive M stars \citep{Lim1996}. \citet{Scuderi1998} observed early type O and B supergiants to make a detailed comparative study of the mass loss evaluated from H$\alpha$ and radio continuum observations. \citet{Guedel1998} and \citet{Gaidos2000} used the Very Large Array (VLA) to search for radio emission of the active, young, solar-type stars $ \rm \pi^{1} \ UMa $, $ \rm \kappa^{1} \ Cet $ and $ \rm \beta \ Com $ at 8.4 GHz. Their observations resulted in $ \rm 3 \sigma $ detection limits of 20 - 30 $ \rm \mu Jy$, which correspond to radio luminosities of $ \rm \sim 10^{12.5} \ erg \ s^{-1} \ Hz^{-1} $ \citep{Villadsen2014}. Early radio observations of EK Dra were recorded in \citet{Guedel1995}, where at minimum the 8.4 GHz flux was $ \rm (34 \pm 11) \ \mu Jy$, and at intermediate levels $ \rm (77 \pm 9) \ \mu Jy $.

To derive an estimate or upper limit for the enhanced young solar wind, we observed radio emission of young, solar-like analogues at the main-sequence with the Karl G. Jansky Very Large Array (VLA) and the Atacama Large Millimeter/Submillimeter Array (ALMA). If we are able to detect free-free radio emission of such winds, their mass loss rates can be calculated. From climate predictions the initial (zero-age main-sequence, ZAMS) solar mass is required to be in the range of 1.03 - 1.07 $ \rm M_{\odot} $ if it were to solve the FYSP \citep{Sackmann2003, Whitmire1995}, suggesting an enhanced early wind mass loss of the order of $ \rm 10^{-12} - 10^{-10} \ M_{\odot} \  yr^{-1} $. In comparison, the present-day solar wind mass loss amounts to $ \rm 2 \times 10^{-14} \ M_{\odot} \  yr^{-1} $ \citep{Feldman1977}. \\

In this paper we focus on the four young solar analogues EK Dra,  $ \rm \pi^{1} \ UMa $, $ \rm \chi^{1}\  Ori, $ and $ \rm \kappa^{1} \ Cet $ using the upgraded sensitivity and resolution of the VLA. In Section \ref{sec:observations}, we briefly describe the observations including a description of our targets. Section \ref{sec:results} contains the results of our detections and upper limits of radio emission. The calculation of the mass loss rates of our star sample will be described in Section \ref{sec:massloss}, where we will also compare our observational results to results from Lyman-$\alpha$ absorption, presented by \citet{Wood2002, Wood2005a}.

\section{Observations}
\label{sec:observations}
\subsection{Target sample}
Our target sample includes the following objects (see also Table \ref{tab:target_sum} summarizing the properties of our stars):

\textit{EK Dra}: This is a G1.5 V star that is considered to be among the most active solar analogues in our neighbourhood, with a distance of 34 pc from the Sun. The average rotation period is 2.68 days. Main properties are reviewed by \citet{Strassmeier1998} and \citet{Messina2003}. \citet{Ribas2005} adopted an age of about 100 Myr for this near-ZAMS star.

\textit{$ \pi^{1} \ UMa $}: This is a young, active G1.5 V solar proxy with a rotation period of about 4.9 days \citep{Messina2003} and a distance of 14.3 pc. In the Sun in Time programme, $ \rm \pi^{1} \ UMa $ is reported to have an age of ~300 Myr \citep{Ribas2005}.

\textit{$ \chi^{1} \ Ori $}: A G1V star with a rotation period of about 5.2 days \citep{Messina2001}, a distance of 8.7 pc and an age of ~300 Myr \citep{Ribas2005}.  The star $ \rm \chi^{1} \ Ori $ is classified as a member of the Ursa Major moving group \citep{King2003}.

\textit{$ \kappa^{1} \ Cet $}: With a spectral type G5 V, it is the coolest star in the sample, with a distance of 9.2 pc from the Sun. \citet{Gaidos2002} determined spectroscopic parameters. The rotation period is reported by \citet{Messina2003} to be about 9.2 days and the age is suggested to be around 650 Myr \citep{Ribas2005, Ribas2010}.

\begin{table*}
 \begin{center}
  \caption{Target characteristics from the Sun in Time programme in \citet{Ribas2005} and \citet{Guedel2007}.}
   \begin{tabular}{lccccccccc}   \hline \hline
    &  &  & d & $ \rm T_{eff} $ & Mass & Radius & $ \rm log \ L_{x} $ & $ \rm P_{rot} $ & Age \\
    Name & HD & Spectral type & (pc) & (K) & ($ \rm M_{\odot}) $ & ($ \rm R_{\odot}) $ & ($ \rm erg \ s^{-1} $)& (days) & (Gyr) \\
    \hline  
    EK Dra & 129333 & G1.5 V & 34.0 & 5870 & 1.06 & 0.95 & 29.93 & 2.68 & 0.1 \\
    $ \rm \pi^{1} \ UMa $ & 72905 & G1.5 V & 14.3 & 5850 & 1.03 & 0.95 & 29.10 & 4.90 & 0.3  \\
    $ \rm \chi^{1} \ Ori ^{a} $ & 39587 & G1 V & 8.7 & 5890 & 1.01 & 0.96  & 28.99 & 5.24 & 0.3 \\
    $ \rm \kappa^{1} \ Cet $ & 20630 & G5 V & 9.2 & 5750 & 1.02 & 0.93 & 28.79 & 9.21 & 0.65 \\
    \hline
  \end{tabular}
  \label{tab:target_sum}
 \end{center}
 \footnotesize{$^{a}$ $ \rm \chi^{1} \ Ori $ has a M-dwarf companion.}
\end{table*}

\subsection{VLA and ALMA}
For the radio measurements we use the Karl G. Jansky VLA, a radio interferometer located in New Mexico near Socorro, operated by the National Radio Astronomy Observatory (NRAO). We use C-band (4 - 8 GHz, 6 cm) and Ku-band (12 - 18 GHz, 2 cm) receivers. The Jansky VLA operates with an increased sensitivity relative to the VLA. The observations were performed in C configuration in sessions in spring/summer 2012 and 2013. The Common Astronomy Software Application (CASA) developed by the NRAO has been used for inspecting, editing (including flagging), calibrating, and imaging the data sets. Flux calibrators were observed at the beginning of each observation for several minutes and the phase calibrators were repeatedly observed together with the targets. An overview and summary of the observations is given in Table \ref{tab:obs_sum}. For the calibration, the raw data needs to be inspected first, which means that bad data due to antenna errors, shadowed antennas, or poor weather conditions need to be flagged, that is removed from the data set. Afterwards, flux, bandpass, and gain calibration steps are applied. We used a pipeline for VLA data\footnote[1]{VLA Calibration Pipeline: https://science.nrao.edu/facilities/vla/data-processing/pipeline} that deals with the flagging and calibration. We used this pipeline, but additional flagging was necessary afterwards.

The Atacama Large Millimeter/submillimeter Array (ALMA), located in the Chajnantor plain of the Chilean Andes, was used to observe in band 3 (with a bandwidth of 84-116 GHz) at 100 GHz in December 2013 within Cycle 1. We got observing time for one of our targets, $ \rm \chi^{1} \ Ori $. For the ALMA data the NRAO staff provided prefabricated scripts together with our data for flagging and calibration. In the meantime, a calibration pipeline for ALMA has been developed as well.\footnote[2]{ALMA Pipeline: http://casa.nrao.edu/casa\_obtaining.shtml.} For our data analysis, we used these pipelines, but some extra flagging and a second run through the pipeline calibration were necessary.

From NRAO's exposure calculator for the VLA, the theoretical noise sensitivity with 2 GHz bandwidth, 27 antennas, and one hour on source is calculated to be around 3.5 $ \rm \mu Jy $ rms in C-band and around 3.8 $ \rm \mu Jy $ rms in Ku-band. These values represent the expected random noise levels for $ \rm \pi^{1} \ UMa $ and $ \rm \kappa^{1} \ Cet $, respectively. Except for $ \rm \pi^{1} \ UMa $ in C-band, the achieved noise levels are in good agreement with the expected values (see Table \ref{tab:obs_sum}). For EK Dra with a bandwidth of 3.5 GHz and 26 antennas, we would expect a noise level of 2.2 $ \rm \mu Jy$ in C-band and 2.9 $ \rm \mu Jy$ in Ku-band, whereas the achieved values are slightly higher. For $ \rm \chi^{1} \ Ori $, 3.5 GHz bandwidth and 26 antennas, the theoretical rms is 1.6 $ \rm \mu Jy $ in C-band, which is in good agreement with the observational noise. In Ku-band the expected noise is around 2.2 $ \rm \mu Jy $, whereas the achieved rms is lower, namely 1.6 $ \rm \mu Jy $. The CLEAN procedure is applied to produce images, where the Clark algorithm with natural weighting was chosen for the setting. Depending on the wavelength and the number of antennas, a cell size of 0.7\arcsec \ and 0.3\arcsec \ for C and Ku-band was used, respectively. \\

\begin{table*}
 \begin{center}
 \caption{Observation summary of our four solar-type targets including the best achieved rms, the clean beam size, and the used phase and flux calibrators. The VLA C-band is defined between 4-8 GHz, Ku-band at 12-18 GHz, and band 3 for ALMA at 100 GHz. Only $ \rm \chi^{1} \ Ori $ was observed with ALMA.}
  \begin{tabular}{lcccccccccccc}   \hline \hline
    Target & Band & Time on &  \#  of & Obs. date & RMS & Clean beam & Phase & Flux \\
    & & source (h) & obs. sets &  & $ \rm (\mu Jy) $ & size & calibrator & calibrator \\
    \hline  
    EK Dra & C & 2 & 1 & Apr 2012 & 3.4 & 4.\arcsec 96 $\times$ 3.\arcsec 43 & J1436+6336 & 3C286 \\
    & Ku & 1.5 & 1 & Apr 2012 & 4 & 2.\arcsec 89 $\times$ 2.\arcsec 26 & J1436+6336 & 3C286 \\
                $ \rm \pi^{1} \ UMa $ & C & 1 & 3 & Jul - Aug 2013 & 7.7 & 5.\arcsec 89 $\times$ 4.\arcsec 50 & J0921+6215 & 3C48 \\
    & Ku & 1 & 4 & Jul 2013 & 2.1 & 2.\arcsec 39 $\times$ 2.\arcsec 01 & J0921+6215 & 3C48 \\
                $ \rm \chi^{1} \ Ori $ & C & 3.75 & 2 & Apr 2012 & 1.8 & 4.\arcsec 34 $\times$ 3.\arcsec 58 & J0559+2353 & 3C147 \\
    & Ku & 2.5 & 6 & Apr - June 2012 & 1.6 & 1.\arcsec 20 $\times$ 0.\arcsec 71 & J0559+2353 & 3C147 \\
    & 3 (ALMA) & 2.5 & 1 & Dec 2013 & 8 & 2.\arcsec 23 $\times$ 2.\arcsec 00 & J0604+2429, & Ganymede \\
    & & & & & & & J0538-4405 \\
    $ \rm \kappa^{1} \ Cet $ & C & 1 & 2 & Jul - Aug 2013 & 3 & 4.\arcsec 07 $\times$ 5.\arcsec 28 & J0339-0146 & 3C48 \\
    & Ku & 1 & 2 & Jul 2013 & 3 & 1.\arcsec 83 $\times$ 2.\arcsec 33 &J0339-0146 & 3C48 \\
    \hline
  \end{tabular}   
  \label{tab:obs_sum}
 \end{center}
\end{table*}

\section{Results}
\label{sec:results}
\subsection{Images}
\begin{figure*}
 \begin{center}
        \subfigure[]{\resizebox{7.55cm}{!}{\includegraphics{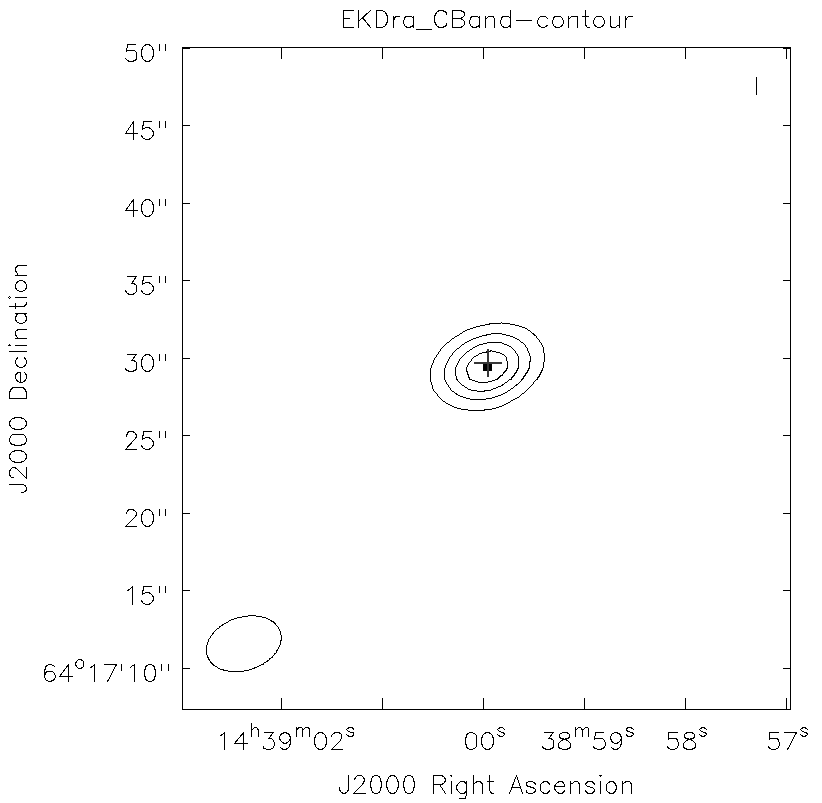}}}
        \subfigure[]{\resizebox{7.55cm}{!}{\includegraphics{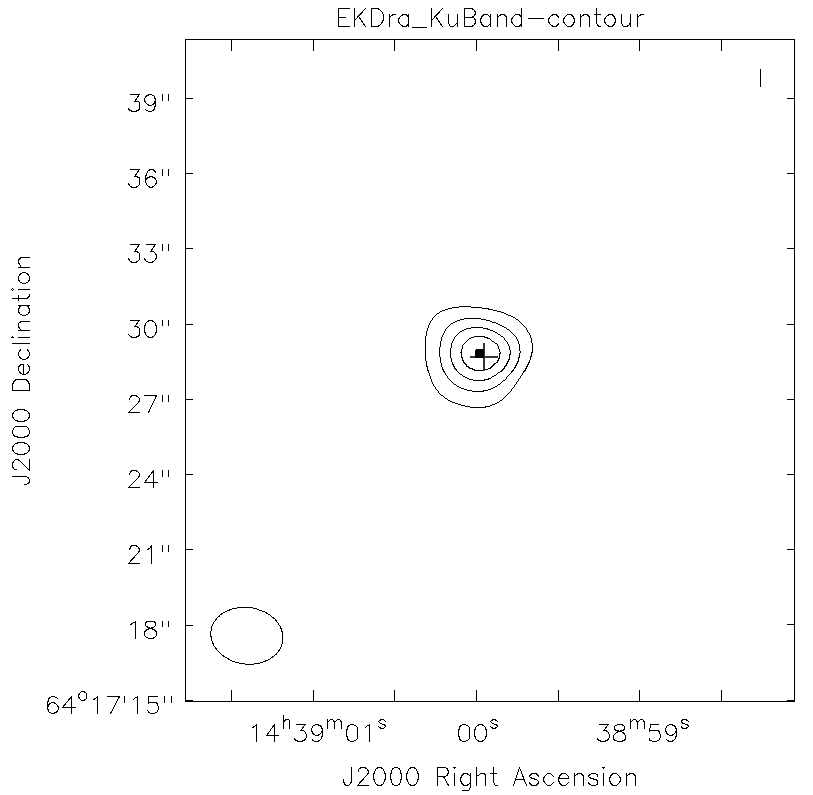}}}
        \caption{Contours of EK Dra in a) C-band with contour levels [0.2, 0.4, 0.6, 0.8] $\cdot$ 576 $\mu$Jy, and b) Ku-band with [0.2, 0.4, 0.6, 0.8] $ \cdot $ 74.5 $\mu$Jy and the beam size in the left corner. The point in the centre gives the coordinates from the Gaussian fit, the cross marks the proper motion corrected Hipparcos position.}
        \label{fig:EK Dra}
\end{center}
\end{figure*}

\begin{figure*}
 \begin{center}
        \subfigure[]{\resizebox{7.0cm}{!}{\includegraphics{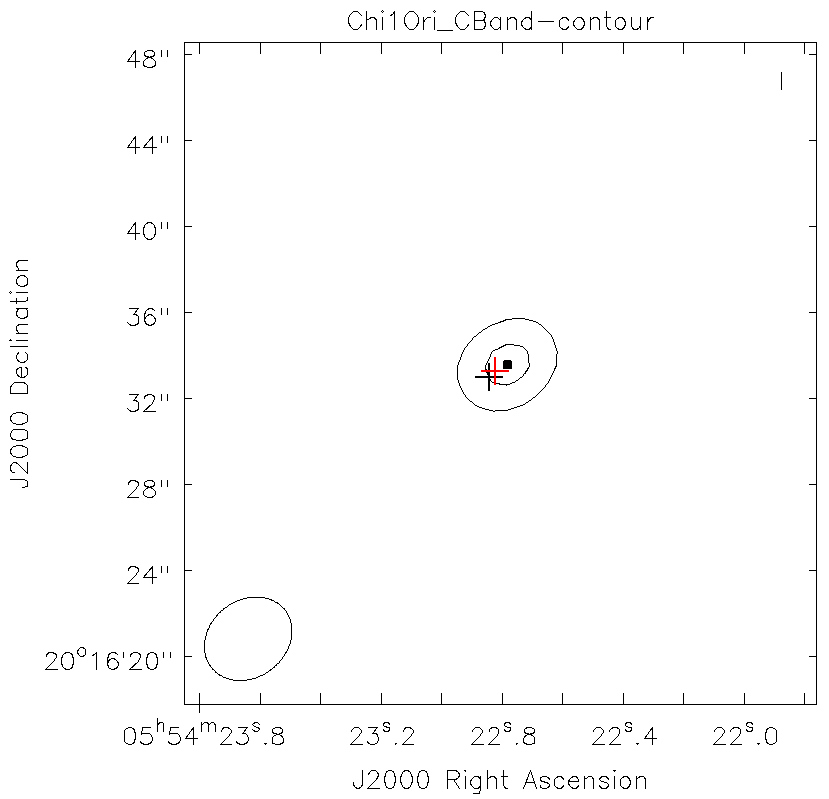}}}
        \subfigure[]{\resizebox{7.2cm}{!}{\includegraphics{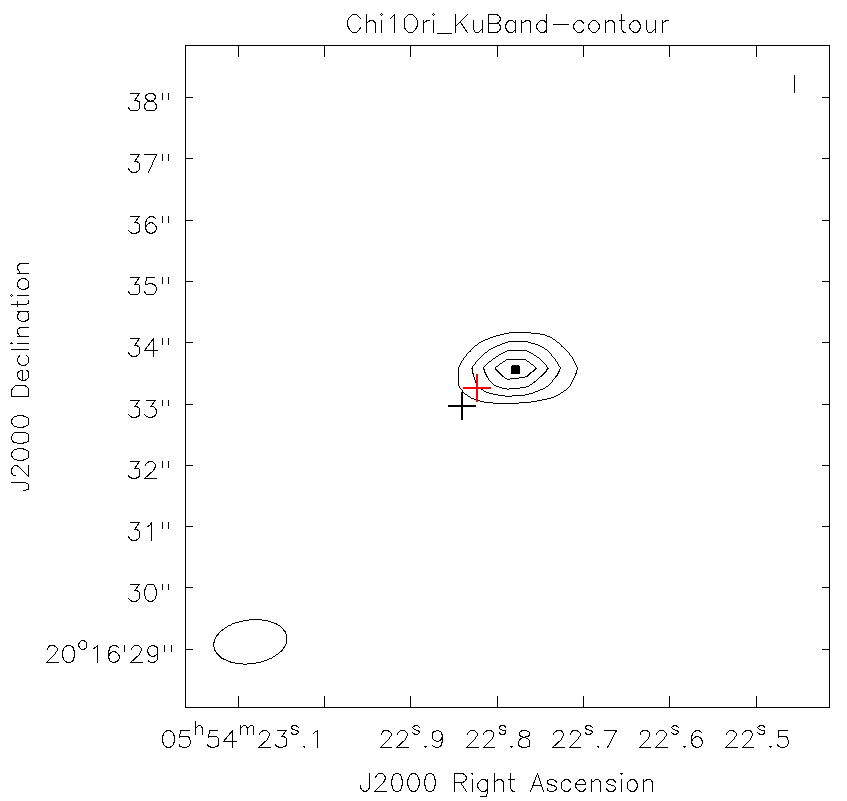}}}
        \subfigure[]{\resizebox{7.4cm}{!}{\includegraphics{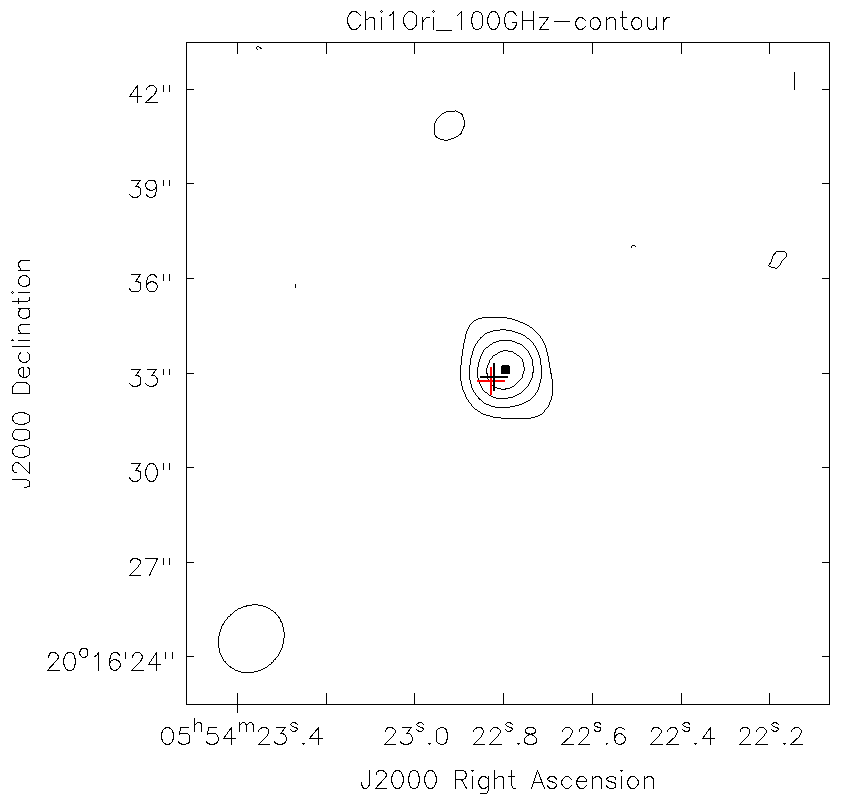}}}
        \caption{$ \rm \chi^{1} \ Ori $ in a) C-band with contour levels [0.2, 0.4, 0.6, 0.8] $ \cdot $ 223 $\mu$Jy,  b) Ku-band with [0.2, 0.4, 0.6, 0.8] $ \cdot $ 106.6 $\mu$Jy, and c) for ALMA at 100 GHz with [0.2, 0.4, 0.6, 0.8] $ \cdot $ 108 $\mu$Jy. The black dot represents the coordinates calculated from the Gaussian fit. The red cross marks the position of the companion $ \rm \chi^{1} \ Ori $ B and the black cross the corrected coordinates of the main target $ \rm \chi^{1} \ Ori $.}
        \label{fig:Chi1Ori}
\end{center}
\end{figure*}

For each target several observation sets are available. To obtain the final images of each target, all data sets in each frequency band are combined to concatenated images which are shown for the detections in Figs. \ref{fig:EK Dra} and \ref{fig:Chi1Ori}. The crosses mark the expected positions of the sources predicted from the \textit{SIMBAD} Astronomical Database\footnote[3]{http://simbad.u-strasbg.fr/simbad/.} corrected for proper motion to the epoch of observation. Minor offsets in right ascension and declination for EK Dra and $ \rm \chi^{1}\  Ori $ occur in our analysis. We note that the field of view in the Ku-band images is much smaller than the C-band images. Several sources can be identified in all images, but only two of the four objects of the sample show a radio detection signal at the expected positions. The targets EK Dra and $ \rm \chi^{1}\  Ori $, shown in Figs. \ref{fig:EK Dra} and \ref{fig:Chi1Ori}, are detected in Stokes I (total intensity), with a total flux of around 100 $ \rm \mu Jy $. 
On the other hand, $ \rm \pi^{1} \ UMa $ and $ \rm \kappa^{1} \ Cet $ (not shown) display non-detections at the expected target positions both in C and Ku-band.

\subsection{Radio emission from the stellar wind}
The free–free spectrum from thermal bremsstrahlung radiation is characterized by a power-law spectral index $ \rm \alpha $, ranging from  $ \rm -0.1 \leq \alpha \leq 2 $, where the flux density is given by $ \rm S_{\nu} \propto \nu^{\alpha} $ at the frequency $ \nu $. The integrated flux densities of the detections are determined by fitting the stellar images by a Gaussian profile. The associated rms values in Stokes I and V (circularly polarized intensity) in both wavelength bands are given in Table \ref{tab:detections}. For those objects for which no detection was observed, the 3$\sigma$ upper limit to the flux density from a source-free background region is estimated. Time series of the sources provide information on the variation of each observation interval. For each object, separate images for each time interval of about five minutes in the Stokes I/V plane are created and hence, the time-dependent flux and the related rms are extracted. Because the observing time is much shorter than the stellar rotation periods, no rotational modulation should be seen. On the other hand, short time variations of a few minutes are indicators for flares.

\begin{table}
 \begin{center}
 \caption{Radio fluxes with their uncertainties in Stokes I and Stokes V, respectively, of the detected objects EK Dra and $ \rm \chi^{1} \ Ori $ and of the non-detections $ \rm \pi^{1} \ UMa $ and $ \rm \kappa^{1} \ Cet $. The upper limits of the fluxes of the non-detections are determined by adopting the 3$\sigma$ background noise as estimation, where  $ \rm \sigma = \sqrt{(rms)^{2}} $.}
  \begin{tabular}{lccccc} \hline \hline
  	Object & \multicolumn{2}{c}{$ \rm S_{\nu} (\mu Jy) $ Stokes I} & \multicolumn{2}{c}{$ \rm S_{\nu} (\mu Jy) $ Stokes V} \\ 	
  	& 6 GHz & 14 GHz & 6 GHz & 14 GHz \\
  	\hline
		EK Dra & 593 $ \pm $ 1.7 & 73 $ \pm $ 2.4  & -22 $ \pm $ 0.8 &  - \\
		$ \rm \pi^{1} \ UMa $ & $\leq$ 23.1 & $\leq$ 6.3 & $\leq$ 8.4 & $\leq$ 6.6 \\
  	$ \rm \chi^{1} \ Ori $ & 110 $ \pm $ 0.7 & 117 $ \pm $ 2.7 & 14 $ \pm $ 0.6 & 12 $ \pm $ 1.1 \\
  	$ \rm \kappa^{1} \ Cet $ & $\leq$ 9 & $\leq$ 9 & $\leq$ 6.9 & $\leq$ 8.7 \\
  	\hline
  \end{tabular}
  \label{tab:detections} 
 \end{center}
\end{table}

The results for the four targets are summarized in the following sections.
\subsubsection{EK Dra}
We obtained a clear detection of EK Dra at $ \rm 14^{h} 38^{m} 59^{s}.96, +64^{\circ} 17^{\prime} 29^{\prime\prime} .49 $. The offsets from the predicted positions in C-band are 0.02s in right ascension and -0.19\arcsec \ in declination, and 0.04s and 0.84\arcsec \ in Ku-band, well within the beam size of 4.\arcsec 96 $\times$ 3.\arcsec 43 and 2.\arcsec 89 $\times$ 2.\arcsec 26, respectively. Because EK Dra is a very active star we expect that the radio emission will include coronal emission \citep{Guedel1995}. The Stokes I radio flux was 593 $ \pm $ 1.7 $ \rm \mu Jy $ with an rms of 3.4 $ \rm \mu Jy$ in C-band. Judging from the light curve, no flare event seems to be present. In Ku-band the radio emission is observed at 73 $ \pm $ 2.4 $ \rm \mu Jy $ with an rms of 4 $ \rm \mu Jy$. The star EK Dra's radio emission cannot be only thermal free-free emission as also argued by \citet{Guedel1995}, given the variability and the high flux level. The polarization degree $ r_{c} = V/I $, which ranges from -1 to 1, is found to vary in the range $r_{c}$ = [-0.088, -0.015] in C-band. In Ku-band our observation does not show any significant non-zero Stokes V flux.
\subsubsection{$ \rm \pi^{1} \ UMa $}
The star $ \rm \pi^{1} \ UMa $, expected at $ \rm 08^{h} 39^{m} 11^{s}.65, +65^{\circ} 01^{\prime} 16^{\prime\prime} .46 $, is a non-detection and was already studied by other authors \citep[e.g.][]{Gaidos2000}. The 3$\sigma $ upper limits of the integrated radio intensities are 23.1 $ \rm \mu Jy $ in C-band and 6.3 $ \rm \mu Jy $ in Ku-band. The C-band intensity limit is high compared to the Ku-band results because, despite heavy flagging and cleaning, the residual of a strong source strongly perturbs our object region and consequently raises the rms. During the observation the fringe pattern directly crossed the expected position of $ \rm \pi^{1} \ UMa $ and caused an increase in the background noise and hence negatively influenced the radio emission estimation for $ \rm \pi^{1} \ UMa $. Therefore, the radio flux density upper limit in C-band is not as useful as desired. On the other hand, the observations of the Ku-band flux density upper limit of 6.3 $ \rm \mu Jy $ are excellent and useful for further analysis and interpretation. The polarization map also shows only noise.
\citet{Gaidos2000} reported a non-detection at the location of $ \rm \pi^{1} \ UMa $ as well. They placed a $2\sigma$ upper limit of 12  $ \rm \mu Jy $ at 3.6 cm (X-band) for the total flux density. Our VLA observations lower these upper limits by a factor of around two.
\subsubsection{$ \rm \chi^{1} \ Ori $}
\begin{figure}
 \begin{center}
        \resizebox{9cm}{!}{\includegraphics{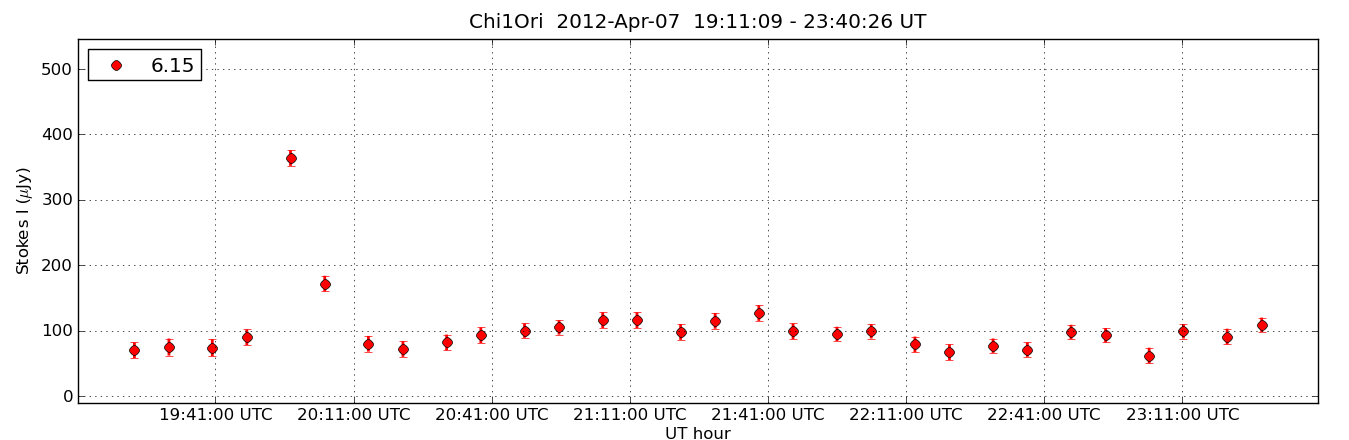}}
        \caption{Time series of $ \rm \chi^{1} \ Ori $ in C-band showing an explosive increase in intensity identified as a flare event at the central frequency of 6.15 GHz. The time interval between the data points is around ten minutes.}
        \label{fig:Chi1Ori_time}
\end{center}
\end{figure}

The star $ \rm \chi^{1} \ Ori $ is located at $ \rm 05^{h} 54^{m} 22^{s}.78, +20^{\circ} 16^{\prime} 33^{\prime\prime} .58 $ in our observations. It shows strong radio emission, seen with offsets in C-band of -0.06s in right ascension and 0.62\arcsec \ in declination, relative to the expected position (cross in Fig. \ref{fig:Chi1Ori}a) using Hipparcos\footnote[4]{http://archive.ast.cam.ac.uk/hipp/hipparcos.html.} measurements  (\citet{vanLeeuwen2007}). In Ku-band the offsets to the observational positions are -0.07s in right ascension and 0.61\arcsec \ in declination. The integrated radio flux densities in Stokes I as given in Table \ref{tab:detections} are 110 $ \pm $ 0.7 $ \rm \mu Jy$ with an rms of 1.8 $ \rm \mu Jy$ in C-band and 117 $ \pm $ 2.7 $ \rm \mu Jy $ and a corresponding rms of 1.6 $ \rm \mu Jy$ in Ku-band. The flux density at 100 GHz measured with ALMA is 103 $ \pm $ 4.9 $ \rm \mu Jy $ in Stokes I. The proper motion corrected offset in right ascension is around -0.03s, in declination it is 0.29\arcsec. The C-band light curve shows a flare that can be clearly identified during an observation interval with a duration of less than 30 minutes (see Fig. \ref{fig:Chi1Ori_time}). The peak reaches a flux density about three times the quiescent level. The occurrence of the flare and the fact that the slope of the spectrum is slightly negative with increasing frequency, suggest that the radio emission of $ \rm \chi^{1} \ Ori $ is not exclusively thermal radio bremsstrahlung from a wind but is dominated by gyrosynchrotron emission from accelerated electrons. The third indication supporting this assumption is a $\approx$ 10\% Stokes V signal (see Table \ref{tab:detections}). The degree of circular polarization $ r_{c} $  is in the range $r_{c}$ = [-0.35, 0.63] with a maximum sigma $ \sigma = $ 0.18 in C-band and $r_{c}$ = [-0.68, 0.72] with $ \sigma $ = 0.49 in Ku-band. For ALMA, no Stokes V measurements were available for Cycle 1 data sets. 
\newline

The images of $ \rm \chi^{1} \ Ori $ show that the corrected coordinates (black crosses in Fig. \ref{fig:Chi1Ori}) do not properly match with the observational positions from the Gaussian fit (black dots). Therefore, we analyzed if the radio signal may come from the M-dwarf companion of $ \rm \chi^{1} \ Ori $ \citep{Han2002, Koenig2002}. To derive the position of $ \rm \chi^{1} \ Ori $ B, the orbit of $ \rm \chi^{1} \ Ori $ has to be corrected first. The orbital parameters are taken from \citet{Han2002} and \citet{Koenig2002}. By correcting the orbit from JD1991.25 to JD2012.4 when our VLA observations took place and by including the correction for proper motion from Hipparcos, the expected coordinates for $ \rm \chi^{1} \ Ori $ are derived. The position of the companion is determined by using the mass ratio between primary and companion, and is displayed by the red cross in the images of Fig. \ref{fig:Chi1Ori} and listed in Table \ref{tab:target_coordinates} (in C-band only). The two components are separated by 0.49\arcsec from each other. Some systematic errors occur from Hipparcos itself, especially because Hipparcos did not recognize the binarity of $ \rm \chi^{1} \ Ori $, and errors in proper motion and possible position errors of the phase calibrator during the observation may contribute to the residual deviation of the detected coordinates. We checked for new Gaia position measurements,\footnote[5]{http://gea.esac.esa.int/archive/.} but unfortunately there is no data available for $ \rm \chi^{1} \ Ori $. If the companion is responsible for the radio emission, which seems likely, we will still use the observational radio emission of  $ \rm \chi^{1} \ Ori $ A or B for our further analysis and mass loss rate calculations considering it to be an upper limit to the thermal wind emission.

\begin{table*}
 \begin{center}
 \caption{Observed and corrected coordinates (proper motion and orbit corrected) as well as the offset between both, for EK Dra, $ \rm \chi^{1} \ Ori, $ and the M-dwarf $ \rm \chi^{1} \ Ori $ B in C-band.}
  \begin{tabular}{lccccccc}   \hline \hline
    & \multicolumn{2}{c}{observed} & \multicolumn{2}{c}{corrected} & \multicolumn{2}{c}{offset} \\
    Name & RA & DEC & RA & DEC & RA & DEC \\
    \hline  
    EK Dra & $ \rm 14^{h} 38^{m} 59^{s}.96$ & $ +64^{\circ} 17^{\prime} 29^{\prime\prime} .49 $ & $ \rm 14^{h} 38^{m} 59^{s}.94 $ & $ +64^{\circ} 17^{\prime} 29^{\prime\prime} .68 $ & -0.02s & -0.19\arcsec \\
    $ \rm \chi^{1} \ Ori $ & $ \rm 05^{h} 54^{m} 22^{s}.78 $ & $ +20^{\circ} 16^{\prime} 33^{\prime\prime} .58 $ & $ \rm 05^{h} 54^{m} 22^{s}.84 $ & $ +20^{\circ} 16^{\prime} 32^{\prime\prime} .96 $ & -0.06s & 0.62\arcsec \\
    $ \rm \chi^{1} \ Ori $ B & $ \rm 05^{h} 54^{m} 22^{s}.78 $ & $ +20^{\circ} 16^{\prime} 33^{\prime\prime} .58 $ & $ \rm 05^{h} 54^{m} 22^{s}.81 $ & $ +20^{\circ} 16^{\prime} 33^{\prime\prime} .24 $ & -0.03s & 0.34\arcsec \\
    \hline
  \end{tabular}
  \label{tab:target_coordinates}
 \end{center}
\end{table*}

\subsubsection{$ \rm \kappa^{1} \ Cet $}
Another non-detection is $ \rm \kappa^{1} \ Cet $ expected at $ \rm 03^{h} 19^{m} 21^{s}.93, +03^{\circ} 22^{\prime} 13^{\prime\prime} .99 $. We therefore report upper limits for the radio emission. Because of the high sensitivity of the VLA, the surrounding noise can be measured at a very low level although a strong source showing up in the C-band image disturbs the field and contributes to the rms even after careful cleaning. The $3\sigma$ rms noise level is used for an upper limit to the radio emission, which is 9 $ \rm \mu Jy $ both in C-band and Ku-band.

\subsection{Chromospheric emission}
We expect that the emission from the stellar chromosphere is small compared to the wind emission. Nevertheless, we estimate the emission from the stellar disk of the star to occur in the chromosphere \citep[see e.g.][for Procyon]{Drake1993}. For an optically thick chromosphere at 10 GHz we assume a temperature of 20 000 K \citep{White2004}. At 100 GHz we expect a lower temperature of typically 10 000 K, although it can be even lower. Furthermore, we assume that the entire surface of the star is covered by chromospheric emission. Using the standard formula for the radio flux from a blackbody with brightness temperature T, the predicted flux density at 100 GHz is

\begin{equation}
 S_{\nu} = \frac{4.94 \cdot 10^{-26}}{d^{2}} \ \ \ \rm erg \ cm^{-2} \ s^{-1} \ Hz^{-1} \ .
\end{equation}

For $ \rm \chi^{1} \ Ori $ at the distance of \textit{d} = 8.7 pc the flux density is 65 $ \mu $Jy for ALMA (100 GHz). Hence, part of the emission observed with ALMA can be of chromospheric origin, but it is probably not the only emission source and cannot explain the detected 100 $\mu$Jy alone. At 10 GHz the expected maximum flux density is 1.3 $\mu$Jy only, and therefore not significant for the VLA detections. For the non-detections, a chromosphere could probably be detected with deeper observations, as in \citet{Villadsen2014}.

\section{Mass loss rates}
\label{sec:massloss}
\subsection{Spherically symmetric winds}
Radio flux density measurements can provide estimates for mass loss rates. The radio free-free flux spectrum for an optically thick, constant velocity, fully ionized isothermal spherical wind is predicted to be of the form \citep{Panagia1975, Wright1975, Olnon1975}:

\begin{equation}
 S_{\nu} = 0.9 \times 10^{11} \ \left(\frac{\dot{M}}{v}\right)^{4/3} T^{0.1} \ \nu^{0.6} \ d^{-2} \ \rm mJy,
\label{eq:flux}
\end{equation}

\noindent
where $ \dot{M} $ is the mass loss rate in $ \rm M_{\odot} \ yr^{-1} $, \textit{T} the temperature of the plasma in K, $ \nu $ the frequency in GHz,{{\changefont{cmr}{m}{it}v}} the wind velocity in $ \rm km \ s^{-1} $ , and \textit{d} the stellar distance in pc. At any frequency one essentially sees emission from gas down to a level where the gas becomes optically thick. \citet{Wright1975}  argue that deviations from $ \rm \alpha_{op} = 0.6 $ (which is the exponent of $ \nu $ in Eq. \ref{eq:flux}) may be caused either by variability due to non-uniform mass loss rates or by an increasing fraction of neutral gas with distance responsible for the radio emission.
Using this formula and assuming a temperature of $ T = 10^{6} \rm \ K $ and an average wind velocity of {{\changefont{cmr}{m}{it}v}} = 400 $ \rm km \ s^{-1} $, the mass loss rate for an (optically thick) wind of $ \rm \pi^{1} \ UMa $ would be $ \dot{M} \leq 1.1 \times 10^{-10} \rm \ M_{\odot} \ yr^{-1} $ for C-band and $ \dot{M} \leq 2.9 \times 10^{-11} \rm \ M_{\odot} \ yr^{-1} $ for Ku-band. The star  $ \rm \kappa^{1} \ Cet $ would show a mass loss rate of $ \dot{M} \leq 2.8 \times 10^{-11} \rm \ M_{\odot} \ yr^{-1} $ for C-band and $ \dot{M} \leq 1.9 \times 10^{-11} \rm \ M_{\odot} \ yr^{-1} $ for Ku-band with the same assumed temperature and velocity profiles. Apart from spherically symmetric (isotropic) winds we will also discuss the possibility of anisotropic, collimated "jet" flows below.

\subsubsection{Radiative transfer equation for non-isothermal winds}
\label{subsec:radtransf}
A point we have to consider is that the temperature in the solar wind (and presumably in winds from other stars) is not constant but decreases with distance \textit{r}. Close to the surface the wind is dense and hot but it cools as it expands. This radial temperature can be roughly described by a $ T \propto r^{-0.5} $ power law \citep{Richardson1995}. Because of this, we wanted to study the case for variable temperature and therefore re-formulated the general radiation transfer equation. 
As described in \citet{Panagia1975}, the intensity $ I_{\nu}(\xi) $  from any line of sight in local thermodynamic equilibrium (LTE) is given by

\begin{equation}
 I_{\nu}(\xi) = B(\nu) (1 - e^{-\tau(\xi)}) \ ,
\end{equation}

\noindent
where $ \xi $ is the distance from the surface of the star out to a boundary of about 200 stellar radii (to ensure that the entire emission region is contained in the calculation volume) measured in the plan perpendicular to the line of sight. A grid for the temperature and density at each grid point was constructed. Emission and absorption were determined for each grid cell at a given distance from the source to create a ring structure with radius $ \xi $ around the source. Moving out to several stellar radii, the contributions from the ring elements are summed up, where the region behind the star is excluded. The optical depth along any line of sight is calculated using:

\begin{equation}
 \tau_{\nu} (s) = \int_{s}^{\infty} n^{2} \ \kappa_{\nu}(T) \ ds \ ,
 \label{eq:tau}
\end{equation}

\noindent
where $ \kappa (\nu) $ is defined as in \citet{Mezger1967}: 

\begin{equation}
 \kappa (\nu) = 8.436 \times 10^{-28} \left[ \frac{\nu}{\rm 10 \ GHz} \right] ^{-2.1} \left[ \frac{T_{e}}{10^{4} \ \rm K} \right] ^{-1.35} \ .
\end{equation}

\begin{figure}
 \begin{center}
        \resizebox{9.1cm}{!}{\includegraphics{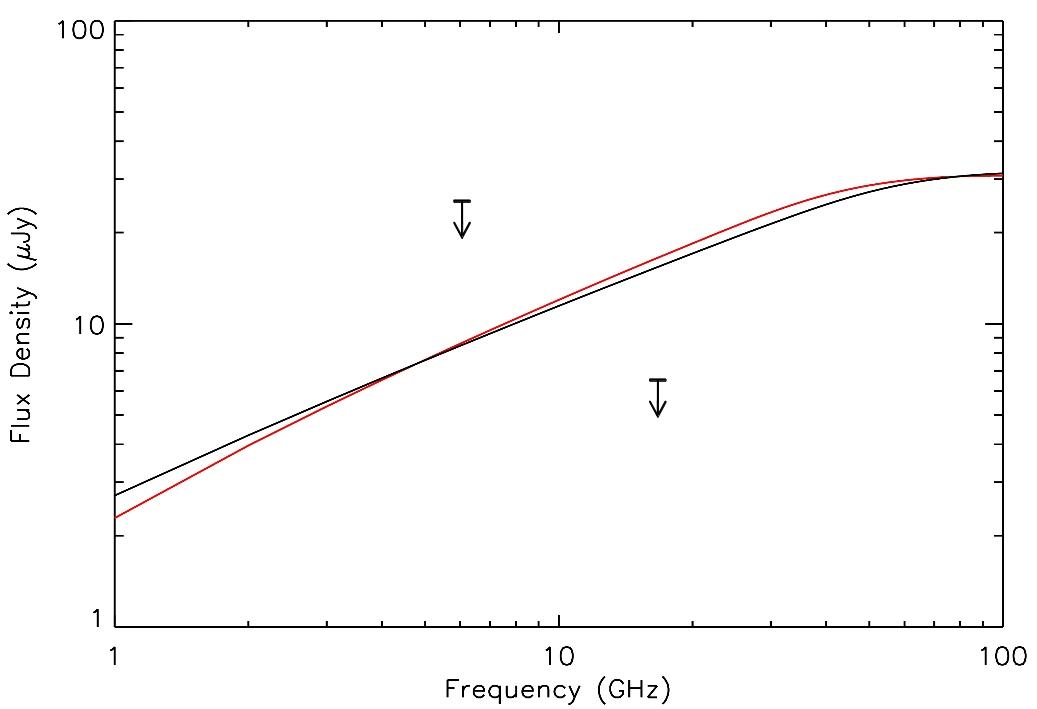}}
        \caption{Example solution (for arbitrary mass loss rate) of the radiative transfer equation assuming a non-constant temperature, shown in red. The black line represents the result for the constant temperature solution. The initial temperature is set to $ T = 10^{6} $ K in both cases. The density is $ n = 2 \times 10^{10} \rm \ cm^{-3} $. The difference in both spectra is not strongly pronounced. The "peak" at lower frequencies in the red curve results from numerical issues. The arrows mark the upper limits of the observational radio fluxes of $ \rm \pi^{1} \ UMa $ in both frequency bands.}
        \label{fig:spec_temp}
\end{center}
\end{figure}

Taking the full geometry into account, we finally obtain a variable temperature transfer equation that can be easily solved numerically. Results are displayed in Fig. \ref{fig:spec_temp}, shown as the red line. The black spectrum displays the solution for a constant temperature. We see that a variable temperature causes minor changes in the steepness of the spectrum which may lead to a slightly higher flux density and may influence the derived mass loss rate. This is probably because of the $n^{2} $ dependence of Eq. \ref{eq:tau} and the strong dependence of the density on distance and hence the temperature, and thus most emission is from very close to the star.
Using the equation of mass continuity $ \dot{M} = 4 \pi r^{2} \rho v, $ the mass loss rate will be approximately 1.1 - 1.6 higher if the temperature is assumed not to be constant. The change in mass loss implied by variations in the temperature is relatively small compared to those due to a change in velocity. The mass loss rate would be enhanced by a factor of about two if the velocity (see Eq. \ref{eq:flux}) changed from {{\changefont{cmr}{m}{it}v}} $ \rm = 400 \ km \ s^{-1} $ to {{\changefont{cmr}{m}{it}v}} $ \rm = 800 \ km \ s^{-1} $.

\subsection{Conical winds}
The stars in our sample are very young and active, hence we investigate anisotropic, collimated winds where the magnetic activity is concentrated at the poles \citep[][and references therein]{Guedel2007}. \citet{Reynolds1986} showed that a well-collimated ionized flow can display a behaviour quite different from that of quasi-spherical flows and they calculated the thermal continuum emission from collimated, ionized winds ("jets") in the presence of gradients in jet width, velocity, ionization, and temperature. 
Because the structure in continuum source spectra contains much information about the flow physics, it is important to get a good frequency coverage of the target sample. The total radio flux of a collimated stellar wind given by \citet{Reynolds1986} is:

\begin{equation}
 S_{\nu} = \int_{y_{0}}^{y_{max}} \left[ \frac{2w(r)}{d^{2}} \right] \left( \frac{a_{j}}{a_{\kappa}} \ T \nu^{2} \right) (1 - e^{-\tau}) \  dy \ .
 \label{eq:total_flux}
\end{equation}

\begin{figure}
 \begin{center}
  \resizebox{9cm}{!}{\includegraphics{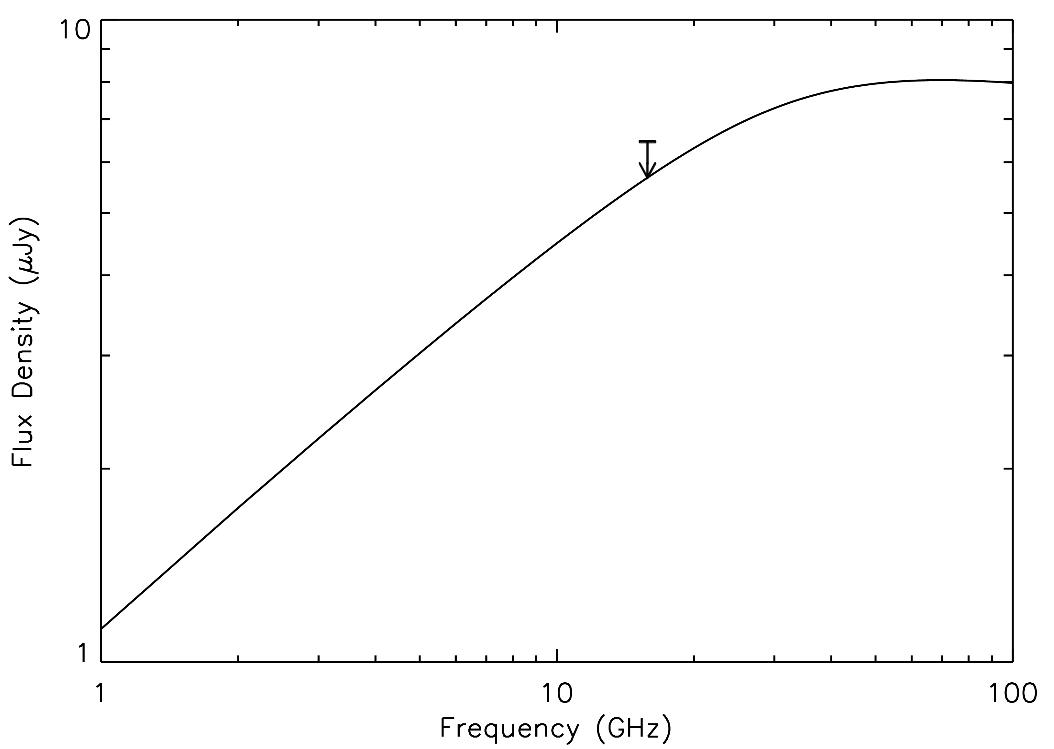}}
  \caption{Upper limits of the total radio flux density for a jet with the properties of $ \rm \pi^{1} \ UMa $ theoretically derived from Eq. \ref{eq:total_flux} with an opening angle of around $ 40^{\circ} $, showing a positive slope of $ \alpha_{op} = 0.6 $ for the optically thick wind and a decrease at the turnover frequency for the optically thin part using the quantities for model A (see Table \ref{tab:values}). The observational upper limit radio flux density of $ \rm \pi^{1} \ UMa $ for Ku-band is shown as a black arrow. The C-band upper limit flux density would lie beyond the plotting range at 23 $ \mu $Jy.}
  \label{fig:integrated_Pi1UMa}
\end{center}
\end{figure}

Here, \textit{y} is defined as \textit{y = r sin i} with \textit{r} being the length of the jet and \textit{i} its inclination (see Fig. 1 in \citet{Reynolds1986}). The half-width of the jet is described with {{\changefont{cmr}{m}{it}w(r)}}, \textit{d} is the distance to the source, \textit{T} the temperature, $ \nu $ the frequency, and $ \tau $ the optical depth based on the wind density, velocity, and temperature. The constants  $ a_{j} = 6.50 \times 10^{-38} $ and $ a_{\kappa} = 0.212 $ link the free-free emission and absorption coefficients: $ j_{\nu}/\kappa_{\nu} = a_{j}/a_{\kappa} \ T \nu^{2} $. The jet half-width, optical depth, temperature, velocity, and density are assumed to vary with $ r/r_{0} $ like power laws with indices $ \epsilon, \ q_{\tau},\ q_{T},\ q_{v,}$ and $ q_{n} $, respectively. The velocity and density indirectly contribute via their indices to the optical depth in Eq. \ref{eq:total_flux}. Different values are assigned to each parameter, depending on the model type \citep{Reynolds1986}, and these quantities are summarized in Table \ref{tab:values}. For example for a constant-velocity, fully ionized, adiabatic jet the exponents are chosen to be $ \epsilon $ = 1, $ q_{n}$ = -2, $ q_{T}$ = -4/3, $ q_{v} $ = 0, and $ q_{\tau} $ = -1.2 (model B, see Table \ref{tab:values}). These variations change the spectral index $ \alpha_{op} $ to 0.83 for a non-isothermal jet instead of $ \alpha_{op} $ = 0.6 for isothermal flows.
Calculating Eq. \ref{eq:total_flux} numerically for the properties of $ \rm \pi^{1} \ UMa $ with $ T = 10^{6} $ K, $ n = 2 \times 10^{10} \rm \ cm^{-3} $, an opening angle of $ 40^{\circ} $ (centred at the pole) and using the standard spherical model quantities (model A), the total flux spectrum is determined and is displayed in Fig. \ref{fig:integrated_Pi1UMa}, where it reveals a positive slope of around $ \alpha_{op} = 0.6 $ for the optically thick wind and a change to $ \alpha = -0.1 $ at high frequencies for the optically thin regime.
\newline

\begin{table}
 \begin{center}
 \caption{Values of $ \epsilon , \ q_{n}, \ q_{T}, \ q_{v}, \ q_{\tau}$ , and $ \alpha_{op} $ for three different models \citep[from][]{Reynolds1986}. Model A represents a "standard" spherical wind, similar to the assumption of \citet{Panagia1975}, but as a conical "jet" with a given opening angle. Model B describes an adiabatic spherical wind and model C an adiabatic collimated wind. The opaque spectral index (i.e. for the optically thick wind) $\alpha_{op}$ is defined as $ \alpha_{op} = 2 + \frac{2.1}{q_{\tau}} (1+ \epsilon + q_{T})$.}
  \begin{tabular}{lcccccccc} \hline \hline
   Model & $ \epsilon $ & $ q_{n} $ & $ q_{T} $ & $ q_{v} $ & $ q_{\tau} $ & $ \alpha_{op} $ & \textit{F} \\
   \hline
   A  & 1 & -2 & 0 & 0 & -3 & 0.6 & 1.5 \\
   B  & 1 & -2 & -$\frac{4}{3} $ & 0 & -1.2 & 0.83 & 3.4 \\
   C  & $ \frac{3}{4} $ & - $ \frac{3}{2} $ & -1 & 0 & -0.9 & 0.25 & 8.0 \\
   \hline
  \end{tabular}
  \label{tab:values}
 \end{center}
\end{table}

To derive upper limits for mass loss rates, different values for velocity and temperature for a standard spherical jet flow, that is  $ \alpha_{op} = 0.6 $, are applied. It is clear that faster but cooler winds lead to stronger mass loss rates. Upper limits for mass loss rates for all three model types, the standard spherical, adiabatic spherical, and adiabatic collimated flow, are calculated using a constant-velocity wind of {{\changefont{cmr}{m}{it}v}} = 400 $ \rm \ km \ s^{-1} $ with the following \citet{Reynolds1986} formula:

\begin{multline}
 \dot{M}_{-6} = 0.938 v_{8} x_{0}^{-1} \left( \frac{\mu}{m_{p}} \right)(S_{mJy} \nu_{10}^{- \alpha_{op}})^{3/4} d_{kpc}^{3/2} \nu_{m10}^{-0.45+3 \alpha_{op} /4} \\ 
 \cdot \theta_{0}^{3/4} T_{4}^{-0.075} (sin \ i)^{-1/4} F^{-3/4} \ ,
\end{multline}

\noindent
where $ \dot{M}_{-6} \equiv \dot{M}/10^{6} \ \rm M_{\odot} \ yr^{-1} $, $ v_{8} \equiv v/10^{8} \ \rm cm \ s^{-1} $, $ \nu_{10} \equiv \nu/10^{10} \ \rm Hz $, $ T_{4} \equiv T/10^{4} \ \rm K $ and

\begin{equation}
 F \equiv \frac{2.1^2}{q_{\tau}(\alpha_{op} - 2)(\alpha_{op} +0.1)} \ .
\end{equation}

\noindent
The frequency $ \nu_{m10} $ is defined as the turnover frequency where the source becomes completely transparent \citep[its definition can be found in Eq. 13 in][]{Reynolds1986} and $ \rm S_{mJy} $ is the observed radio flux of our objects in mJy.
Table \ref{tab:mass_sum} summarizes the maximum mass loss rates for $ \rm \pi^{1} \ UMa $ and $ \rm \kappa^{1} \ Cet $ at 6 GHz and 14 GHz for all three models by changing the model parameters, a temperature of $ T = 10^{6} $ K, a velocity of {{\changefont{cmr}{m}{it}v}} $ \rm = 400 \ km \ s^{-1} $ , and an opening angle of $ 40^{\circ} $.

\begin{table*}
 \begin{center}
 \caption{Upper limits mass loss rates for $ \rm \pi^{1} \ UMa $ and $ \rm \kappa^{1} \ Cet $ for three model types with different calculated $ \alpha_{op} $ with constant velocity of {{\changefont{cmr}{m}{it}v}} $ \rm  = 400 \ km \ s^{-1} $ but varying temperature and density. We see that a spectral index of $ \alpha $ = 0.6 results in the largest mass loss rates.} 
  \begin{tabular}{lcccc} \hline \hline
    & model A ($ \alpha $ = 0.6) & model B ($ \alpha $ = 0.83) & model C ($ \alpha $ = 0.25) \\
    & \multicolumn{3}{c}{ $ \dot{M} \rm \ (M_{\odot} \ yr^{-1}) $} \\
   \hline
   $ \rm \pi^{1} \ UMa $ (C-band) & $ \leq 1.9 \times  10^{-11} $ & $ \leq 1.5 \times  10^{-11} $ & $ \leq 3.1 \times  10^{-12} $\\
   $ \rm \pi^{1} \ UMa $ (Ku-band) &  $ \leq 5.0 \times  10^{-12} $ & $ \leq 3.4 \times  10^{-12} $ & $ \leq 1.0 \times  10^{-12} $\\
   $ \rm \kappa^{1} \ Cet $ (C-band) & $ \leq 5.1 \times  10^{-12} $ & $ \leq 4.0 \times  10^{-12} $ & $ \leq 8.0 \times  10^{-13} $\\
   $ \rm \kappa^{1} \ Cet $ (Ku-band) & $ \leq 3.5 \times  10^{-12} $ & $ \leq 2.4 \times  10^{-12} $ & $ \leq 6.9 \times  10^{-13} $ \\
   \hline
  \end{tabular}
  \label{tab:mass_sum}
 \end{center}
\end{table*}

If we assume that the mass loss rates are a function of the opening angle, they increase with increasing opening angle. For example, the mass loss rate with an opening angle of $ 20^{\circ} $ is $ \dot{M} \leq 3.0 \times  10^{-12} \rm \ M_{\odot} \ yr^{-1} $ for $ \rm \pi^{1} \ UMa $ for Ku-band. Enlarging the angle to $ 60^{\circ} $ the mass loss rate increases to $ \dot{M} \leq 6.7 \times  10^{-12} \rm \ M_{\odot} \ yr^{-1} $. A higher velocity of {{\changefont{cmr}{m}{it}v}} $ \rm = 800 \ km \ s^{-1} $ would raise the mass loss rates by a factor of two. Although we are not able to detect any radio emission signal for $ \rm \pi^{1} \ UMa $ and $ \rm \kappa^{1} \ Cet $, we can thus give meaningful upper limits to the mass loss rates of these young stars within a range of reasonable wind opening angles and wind temperatures.

As already mentioned, additional coronal, partly flaring radio emission for EK Dra and $ \rm \chi^{1} \ Ori $ was detected and identified as non-thermal emission, but we can nevertheless provide meaningful upper limits by adopting the detected non-thermal flux densities as upper limits to the thermal emission. We calculate the maximum mass loss of both stars for a spherically symmetric and a conical wind, as done for the non-detections. These mass loss rates for EK Dra and $ \rm \chi^{1} \ Ori $ in both frequency bands are summarized in Table \ref{tab:mass_detections}.

\begin{table*}
 \begin{center}
 \caption{Upper limits of mass loss rates of EK Dra and $ \rm \chi^{1} \ Ori $ determined for spherically symmetric winds and conical jet flows. All calculations are done in both frequency bands (C and Ku).}
  \begin{tabular}{lccc} \hline \hline
        & Spherically symmetric wind & Conical jet \\
        & \multicolumn{2}{c}{ $ \dot{M} \rm \ (M_{\odot} \ yr^{-1}) $} \\
        \hline
        EK Dra (C-band) & $ 4.6 \times 10^{-9} $ & $ 2.3 \times 10^{-9} $ \\
        EK Dra (Ku-band) & $ 6.5 \times 10^{-10} $ & $ 4.7 \times 10^{-11} $ \\
        $ \rm \chi^{1} \ Ori $ (C-band) & $ 1.7 \times 10^{-10} $ & $ 8.3 \times 10^{-11} $ \\
        $ \rm \chi^{1} \ Ori $ (Ku-band) & $ 1.2 \times 10^{-10} $ & $ 8.7 \times 10^{-11} $ \\
        \hline
  \end{tabular}
  \label{tab:mass_detections} 
 \end{center}
\end{table*}

\subsection{Absorption of the wind due to flares}
\label{subsec:Flares}
The presence of flares and polarized emission on EK Dra and $ \rm \chi^{1} \ Ori $ imply that any radio contribution from winds must be significantly lower than the detected radiation. The non-thermal and flare emission originate close to the surface of the star. The fact that it is detectable implies that the stellar wind is optically thin to this radiation. An assessment for the maximum mass loss possible for an optically thin wind was suggested in \citet{Lim1996}. A stronger wind would completely absorb the observed radiation from coronal radio flares. The radius at which a spherically symmetric wind becomes optically thick at a given frequency $ \nu $ can be derived from the expression \citep{Lim1996}:

\begin{multline}
 \frac{R(\nu)}{R_\odot} \approx 6 \left( \frac{\nu}{10 \ \rm GHz} \right)^{-2/3} \left( \frac{T}{10^{4} \ \rm K} \right)^{-1/2} \\
 \times \left( \frac{\dot{M}}{10^{-10} \ \rm M_\odot \ yr^{-1}} \right)^{2/3} \left( \frac{v_\infty}{300 \ \rm km \ s^{-1}} \right)^{-2/3} .
 \label{eq:lim_eq}
\end{multline}

Because the non-thermal emission from the star must originate above the optically thick surface at the observing frequency to be detectable, we set R($\nu$) to $ R_{\ast} $.
Assuming the terminal velocity to be {{\changefont{cmr}{m}{it}$v_{\infty}$}} = 400 $\rm km \ s^{-1} $ and the temperature $ T = 10^{6}$ K, and solving Eq. \ref{eq:lim_eq} for $ \dot{M} $,  we find a maximum wind mass loss rate of $ \dot{M} \rm \leq 1.3 \times 10^{-10} \ M_{\odot} \ yr^{-1} $ for C-band and $ \dot{M} \rm \leq 6.9 \times 10^{-10} \ M_{\odot} \ yr^{-1} $ for Ku-band for EK Dra. With the same velocity and temperature profiles for $ \rm \chi^{1} \ Ori $, a wind with $ \dot{M} \rm \leq 1.3 \times 10^{-10} \ M_{\odot} \ yr^{-1} $ for C-band at 6 GHz and $\dot{M} \rm \leq 7.2 \times 10^{-10} \ M_{\odot} \ yr^{-1} $ for Ku-band at 14 GHz is derived. Comparing these mass loss rates to those derived for a spherically symmetric and a conical wind, respectively,  as given in Table \ref{tab:mass_detections}, we see that the estimates are similar. We keep the conical wind mass loss results as upper limits for EK Dra and $ \rm \chi^{1} \ Ori $.

\subsection{Rotational evolution}
\label{subsec:rot_evolution}
As magnetized stellar winds remove angular momentum from their host stars and therefore force stars to spin down \citep{Weber1967, Skumanich1972, Kraft1967} and cause a decrease in rotation rate and magnetic activity as they age \citep{Guedel1997, Vidotto2014}, it is essential to consider rotational evolution when determining mass loss rates of young, active stars. Several solar wind models \citep[e.g.][]{vanderHolst2007, Zieger2008, Jacobs2011} and rotational evolution models \citep[e.g.][]{Cranmer2011, Gallet2015} have been developed. \citet{Johnstone2015a} developed a solar wind model to estimate the properties of stellar winds for low-mass main-sequence stars between masses of 0.4 $ M_{\odot} $ and 1.1 $ M_{\odot} $ at a range of distances from the star based on stellar spin-down and angular momentum loss in a magnetized wind. They used 1D thermal pressure-driven hydrodynamic wind models using the Verstile Advection Code \citep{Toth1996} and in-situ measurements of the solar wind. The stellar mass loss rate can then be calculated with

\begin{equation}
 \dot{M} = \dot{M_\odot} R^{2} \Omega^{1.33} M^{-3.36} \ ,
\end{equation}

\noindent
where all quantities are in solar units with the Carrington rotation rate of $ \Omega_{\odot} = 2.67 \times 10^{-6} \rm \ rad \ s^{-1} $. Graphically, this relation is shown in Fig. 10 in \citet{Johnstone2015b}. Applying this formula to our four objects we are able to calculate their mass loss rates considering their rotational evolution, shown as red filled circles in Fig. \ref{fig:massloss_evolution}. These values follow a $ \dot{M} \propto t^{-0.75} $ relation \citep{Johnstone2015b} and are about two orders of magnitude lower than our upper limits displayed as arrows, but we emphasize that these results are indirect inferences from models.

\subsection{Early mass loss of the Sun}
\begin{figure}
 \begin{center}
        \resizebox{9.1cm}{!}{\includegraphics{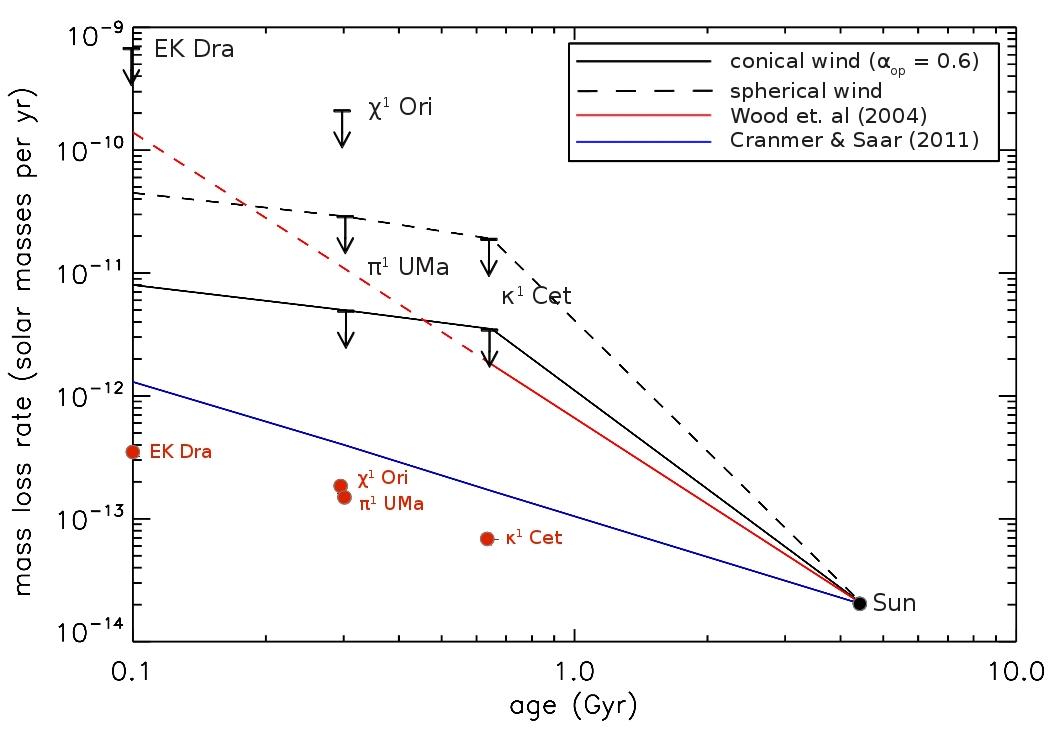}}
        \caption{Mass loss evolution described by the non-detections $ \rm \pi^{1} \ UMa $ and $ \rm \kappa^{1} \ Cet $ at 0.3 and 0.65 Gyr, respectively, and the present Sun at 4.5 Gyr, shown as a black solid line assuming a conical wind with $\alpha_{op} = 0.6$ (model A, solid line). The mass loss rates of all sources including EK Dra and $ \rm \chi^{1} \ Ori $ are determined from their observational flux densities in Ku-band and opening angles of $ 40^{\circ} $. For models B and C, this evolution would lie below the solid line implying a lower upper limit of the solar mass. The black dashed line shows the evolution assuming a spherically symmetric wind.  For EK Dra and $ \rm \chi^{1} \ Ori $ only the mass loss rates of spherically symmetric winds are shown, whereas for $ \rm \pi^{1} \ UMa $ and $ \rm \kappa^{1} \ Cet $, mass loss rates of symmetric and conical winds, respectively, are shown. The red circles are the mass loss rate estimates using rotational evolution model calculations described in \citet{Johnstone2015b}. The red line shows the result of \citet{Wood2004}, where the dashed line indicates the age region where the power law fails. \citet{Cranmer2011} also estimate a mass loss rate versus time; this fit is shown as the blue solid line.}
        \label{fig:massloss_evolution}
\end{center}
\end{figure}

How did the solar wind evolve over time? For a simple evaluation of the total early solar mass loss, power laws are placed through the sample of young, solar-type stars observed in this study. The upper limits of the mass loss rate for the conical wind ($ \alpha $ = 0.6) of the two non-detections of $ \rm \pi^{1} \ UMa $ and $ \rm \kappa^{1} \ Cet $ in Ku-band and the solar wind mass loss rate are used to define a piecewise power law through the sample. These relationships are shown in Fig. \ref{fig:massloss_evolution}. Additionally, the mass loss rates from rotational evolution are marked as red circles. Although EK Dra and $ \rm \chi^{1} \ Ori $ are marked in the plot, they are not used for the evaluation of the power laws, since these $ \dot{M} $ are estimated based on the detected non-thermal radiation. The power laws are extrapolated from 0.3 Gyr down to the age of 0.1 Gyr. First, we apply our results to spherically symmetric winds with the corresponding power laws, which give an upper limit to the solar mass loss of 2.02 \% after the integration from 100 Myr to 4.5 Gyr, resulting in an initial solar mass of 1.02 $ M_{\odot} $. Conical winds (using $\alpha_{op} = 0.6 $) follow similar power laws as shown in Fig. \ref{fig:massloss_evolution}:

\begin{equation}
 \dot{M} \propto t^{-0.46} \ \ \ \ \rm from \ 0.1 \ to \ 0.65 \ Gyr
\end{equation}
\begin{equation}
 \dot{M} \propto t^{-2.66} \ \ \ \ \ \ \ \rm from \ 0.65 \ to \ 4.5 \ Gyr.
\end{equation}

\citet{Cranmer2011} also estimate a mass loss rate versus time resulting in a power law index of -1.1, lying below our mass loss upper limits but above the model calculations for rotational evolution, shown as the blue solid line in Fig. \ref{fig:massloss_evolution}. The relation of \citet{Wood2004} as given in Eq. \ref{eq:wood_flux} is shown as the red line in the figure. The dashed part of the line marks the age region where the relation fails for most of young and active stars. Furthermore, \citet{Alvarado2016} find a weaker power law relation of mass loss rate versus age based on magnetohydrodynamics (MHD) simulations, resulting in $ \dot{M} \propto t^{-1.37} $.

After the integration in time from 100 Myr to 4.5 Gyr, the total mass is in our case for the above given power laws at most 0.4 \% ($ \rm \alpha_{op} = 0.6 $) higher than at present, depending on the model for the spectral index $ \alpha_{op} $, resulting in a solar mass of 1.004 $ M_{\odot} $ only. Considering the theoretical model calculation for rotational evolution, the solar mass would be even lower at 1.0002 $ M_{\odot} $.
The boundaries necessary for solving the FYSP are at 3\% to 7\% total mass loss, required to keep liquid water on early Mars and to control and avoid the runaway greenhouse effect on Earth at early stages up to a few 100 Myr \citep{Whitmire1995}. Our limits for the spherically symmetric and conical wind models are definitely below the 3\% boundary and therefore imply that the faint young Sun problem cannot be solved by assuming increased wind mass loss rates and therefore a higher mass for the young Sun.

\section{Summary and Discussion}
In this study, we analyzed four young, solar-type stars on the main-sequence of different ages, which are part of the Sun in Time programme to study the decline of magnetic activity and wind mass loss in solar analogues. For the analysis, observations of the VLA at 2 cm and 6 cm wavelength and ALMA at 100 GHz are used, aiming to detect thermal radio emission, that is free-free radio bremsstrahlung, which is indicative of the existence of a stellar wind. The well-studied analogues of the Sun cover the young evolutionary stages on the main-sequence. Our sample of four stars results in two detections: EK Dra and $ \rm \chi^{1} \ Ori $; and two non-detections: $ \rm \pi^{1} \ UMa $ and $ \rm \kappa^{1} \ Cet $. For both detections we can conclude that the radio emission is not thermal bremsstrahlung alone, but consists of additional coronal radio emission in the form of non-thermal, partly flaring emission. Indicators for that assumption are a negative slope of the radio spectrum, the presence of flares seen in the light curves, and the presence of circular polarization. Furthermore, we have argued that instead of $ \rm \chi^{1} \ Ori $, we have actually detected its M-dwarf companion.
For the non-detections $ \rm \pi^{1} \ UMa $ and $ \rm \kappa^{1} \ Cet $, we can estimate their maximum wind radio emission flux densities by placing the $ 3 \sigma $ rms value as upper limits. We have to clearly state that we cannot rule out other contributing but also undetected emission processes in these sources.

The estimated radio emissions are used to derive upper limits to mass loss rates for the observed targets. Mass loss rates are important quantities for the study of the evolution of young stars including the Sun. They could possibly result in an explanation and solution for the problem of the famous FYSP. Furthermore, the evolution of mass loss rates of the young Sun leads to essential information for the formation and evolution of the atmospheres of the early Earth and other planetary atmospheres. 
We estimate mass loss rates for all targets for spherically symmetric and anisotropic collimated winds. We note that any additional neutral wind component would increase the mass loss rate, but such winds would not be detected by our methods. However, the solar wind is essentially fully ionized, so we assume the same for solar analogues.
\newline

We applied three different model types (standard spherical, adiabatic spherical, and adiabatic collimated) for the mass loss rate calculation by changing the different parameter quantities. If we vary the velocity and the temperature, we see that the mass loss rate increases for a cooler and faster wind.
 
The star EK Dra's mass loss is estimated to be $ \dot{M} \leq 1.3 \times 10^{-10} \rm \ M_{\odot} \ yr^{-1} $ for C-band and $ \dot{M} \leq 6.9 \times 10^{-10} \rm \ M_{\odot} \ yr^{-1} $ for Ku-band. The star $ \rm \chi^{1} \ Ori $ shows a mass loss rate of $ \dot{M} \leq 1.3 \times 10^{-10} \rm \ M_{\odot} \ yr^{-1} $ for C-band at 6 GHz and $ \dot{M} \leq 7.2 \times 10^{-10} \rm \ M_{\odot} \ yr^{-1} $ for Ku-band at 14 GHz. Here, we assume that non-thermal emission from coronal radio flares contributes to the total mass loss following \citet{Lim1996}, implying that these features originate close to the stellar surface propagating through an optically thin wind. Mass loss rates from the spherically symmetric wind and conical wind calculations are similar to these results.

For $ \rm \pi^{1} \ UMa $ the mass loss rate is derived to be $ \dot{M} \leq 1.9 \times  10^{-11} \rm \ M_{\odot} \ yr^{-1} $ for C-band and $ \dot{M} \leq 5 \times 10^{-12} \rm \ M_{\odot} \ yr^{-1} $ for Ku-band for a jet-like wind with an opening angle of $ 40^{\circ} $. For $ \rm \kappa^{1} \ Cet $ the determined mass loss rate for a collimated wind is $ \dot{M} \leq 5.1 \times 10^{-11} \rm \ M_{\odot} \ yr^{-1} $ for C-band and  $ \dot{M} \leq 3.5 \times 10^{-12} \rm \ M_{\odot} \ yr^{-1} $ for Ku-band.

The resulting maximum mass loss rate of $ \rm \pi^{1} \ UMa $ in Ku-band (with $ \alpha_{op} = 0.6 $) is about 250 times stronger than the present day solar mass loss rate of $ \dot{M} = 2 \times 10^{-14} \rm \ M_{\odot} \ yr^{-1} $. \citet{Wood2014} studied the stellar wind and mass loss of $ \rm \pi^{1} \ UMa $ using Ly-$\alpha$ observations. With hydrodynamic models for the astrosphere to infer the stellar wind strength, the study of \citet{Wood2014} results in a wind for $ \rm \pi^{1} \ UMa $ only half as strong as the solar wind. From their research, \citet{Wood2014} concluded that the Sun and solar-like stars do not experience particularly strong coronal winds in their past.
\citet{Drake2013} studied coronal mass ejections in connection to stellar winds, where the authors found that coronal mass ejection (CME) induced mass loss rates can amount to several percent of the steady wind rate. Their estimation for a CME mass loss rate for $ \rm \pi^{1} \ UMa $ implies $ \dot{M} \sim 3 \times 10^{-12} \rm \ M_{\odot} \ yr^{-1} $, comparable with our upper limits. For $ \rm \kappa^{1} \ Cet $ the mass loss rate in the \citet{Drake2013} study is similar. We see that the measurements of \citet{Wood2014} of $ \dot{M} = 0.5 \ \dot{M}_{\odot} $ are much lower than the $ \dot{M} = 150 \ \dot{M}_{\odot} $ predictions of \citet{Drake2013} and our observational upper limit. The rotation-wind model by \citet{Johnstone2015b} in fact also requires a wind mass loss rate significantly above the value suggested by \citet{Wood2014} to explain the observed spin-down rate for solar-like stars in this age range.
\newline

Finally, the maximum total solar mass for the young Sun was derived for three cases: spherically symmetric winds, conical jet flows, and rotational evolution models. The results are quite different: 1) the mass loss rate of spherically symmetric winds indicates a total maximum mass of 1.02 $ \rm M_{\odot} $, 2) conical winds lead to a total mass of 1.004 $ \rm M_{\odot} $ , and 3) the rotational evolution model suggests an initial solar mass of only 1.0002 $ \rm M_{\odot} $ at an age of about 100 Myr.

\section{Conclusion}
If the FYSP is to be solved with a larger initial solar mass, Earth and Mars climate constraints require the solar mass to be in the range of 1.03 - 1.07 $ \rm M_{\odot} $ near the zero-age main-sequence, requiring an enhanced early wind mass loss rate of order $ \rm 10^{-12}-10^{-10} \ M_{\odot} \ yr^{-1} $. Our results for mass loss rates derived with radio observations of solar analogues indicate an early solar mass of at most $ \rm 1.02 \ M_{\odot} $ assuming spherically symmetric winds. This is not sufficient to solve the faint young Sun paradox. It appears that other explanations such as higher concentrations of greenhouse gases and aerosols \citep[e.g.][]{Sagan1972, Kasting1993}, a lower global albedo, either through less cloud coverage \citep[e.g.][]{Shaviv2003}, and/or a smaller continental land mass \citep{Rosing2010} are required.

\begin{acknowledgement}
We thank the referee, Jeffrey Linsky, for very helpful comments that improved the paper. This research has made use of the SIMBAD database, operated at CDS, Strasbourg, France. B.F and M.G. acknowledge the support of the FWF "Nationales Forschungsnetzwerk" project S116601-N16 "Pathways to Habitability: From Disks to Active Stars, Planets and Life" and the related FWF NFN subproject S116604-N16 "Radiation and Wind Evolution from the T Tauri Phase to ZAMS and Beyond". Financial support of this project by the University of Vienna is also acknowledged. This publication is supported by the Austrian Science Fund (FWF).
The National Radio Astronomy Observatory is a facility of the National Science Foundation operated under cooperative agreement by Associated Universities, Inc. This paper makes use of the following ALMA data: $ \rm ADS/JAO.ALMA\#2011.0.01234.S $. ALMA is a partnership of ESO (representing its member states), NSF (USA) and NINS (Japan), together with NRC (Canada), NSC and ASIAA (Taiwan), and KASI (Republic of Korea), in cooperation with the Republic of Chile. The Joint ALMA Observatory is operated by ESO, AUI/NRAO and NAOJ. 
\end{acknowledgement}

\bibliographystyle{aa}
\bibliography{A&AFichtinger2014}

\end{document}